# Synthesis and Bulk Properties of Oxychloride Superconductor $Ca_{2-x}Na_xCuO_2Cl_2$

N. D. Zhigadlo[(1)], J. Karpinski[(1)], S. Weyeneth[(2)], R. Khasanov[(2)], S. Katrych[(2)], P. Wägli[(1)] and H. Keller[(2)]

(1) Laboratory for Solid State Physics, ETH Zurich, 8093 Zurich, Switzerland
e-mail: zhigadlo@phys.ethz.ch

(2) Physik-Institut der Universität Zürich, 8057 Zürich, Switzerland

*Abstract* - **A series of polycrystalline samples and submillimeter size single crystals of a cuprate oxychloride $Ca_{2-x}Na_xCuO_2Cl_2$ (Na-CCOC) with values of Na content ranging from underdoped to optimally doped regions were synthesized at pressure of 30-55 kbar and temperature of 1250-1700 ºC. A systematic variation of the transition temperature $T_c$ with a maximum value of 29 K for $x \approx 0.20$ has been found as a function of Na content. In order to check the role of the apical oxygen for high-temperature superconductivity, we performed muon-spin rotation and magnetization studies of the in-plane magnetic penetration depth $\lambda_{ab}$ for $Ca_{2-x}Na_xCuO_2Cl_2$ samples with $x \approx 0.11$, 0.12, 0.15, 0.18, and 0.19. The absolute value of the in-plane magnetic penetration depth at $T=0$ was found to increase with decreasing doping from $\lambda_{ab}(0)=316(19)$ nm for the $x \approx 0.19$ sample to $\lambda_{ab}(0)=430(26)$ nm for the $x \approx 0.11$ one. Based on a comparison of the present Na-CCOC data with the data of $La_{2-x}Sr_xCuO_4$ cuprate superconductors, it is concluded that replacing of apical oxygen by chlorine decreases the coupling between the superconducting $CuO_2$ planes, leading to an enhancement of the two-dimensional properties of Na-CCOC. The torque studies implies that the anisotropy coefficient $\gamma=84$ of $Ca_{1.82}Na_{0.18}CuO_2Cl_2$ single crystals is much more enhanced compared to the structurally related $La_{1.82}Sr_{0.18}CuO_4$ where for same doping $\gamma$ is much lower, i.e. $\gamma \approx 11$.**

## I. INTRODUCTION

$Ca_{2-x}Na_xCuO_2Cl_2$ (Na-CCOC) is a structural analogue to the cuprate superconductor $La_{2-x}Sr_xCuO_4$ (La214) with Cl atoms replacing oxygen on the apical sites [1]. Na-CCOC is an ideal model system for studying of the electronic state of doped $CuO_2$ planes and for establishing the role of the apical sites in high-temperature superconductivity [2]. Despite of similarity to other cuprate superconductors properties of the Na-CCOC compound are not well characterized due to difficulties in obtaining single crystals. Single crystals can be grown only at high pressure and they are very hygroscopic [3, 4]. Investigation of intrinsic superconducting properties on single crystals might help to clarify the origin of similarities and differences between cuprates with oxygen and halogen atoms on the apical sites. With the aim of growing large crystals of Na-CCOC, suitable for physical measurements, we carried out a systematic investigation of the parameters controlling the growth of crystals, including temperature, pressure, composition, reaction time and heating/cooling rate. Typical crystals having several hundreds μm in linear sizes were reproducibly obtained. Based on comparison of results of our muon-spin rotation (μSR), low-field magnetization and torque studies of $Ca_{2-x}Na_xCuO_2Cl_2$ with analogous results for structurally related $La_{2-x}Sr_xCuO_4$ it is concluded that replacing apical oxygen by chlorine decreases the coupling between the superconducting $CuO_2$ planes leading to an enhancement of the two-dimensional properties of Na-CCOC.

## II. EXPERIMENTAL DETAILS

The nonsuperconducting parent compound $Ca_2CuO_2Cl_2$ was synthesized by a solid-state reaction of $Ca_2CuO_3$, $CuO$, and $CaCl_2$. The powder mixture was pressed into a pellet and annealed at 750 °C in argon flow with several intermediate grindings under ambient pressure. The resulting $Ca_2CuO_2Cl_2$ was then well mixed with $NaClO_4$ and $NaCl$ in a dry box and sealed in Pt cylindrical capsules of 6.8-8 mm internal diameter and 7-9 mm length. The $NaClO_4$ decomposes and produces oxygen pressure at high temperature, and then it work as a flux. High-pressure and high-temperature experiments were performed in a cubic anvil and opposed anvil-type high-pressure devices (Fig. 1). The cubic anvil-type device is shown in left panel of Fig. 1. A set of steel parts transmits force through six tungsten carbide pistons to the sample volume in a quasi isostatic way. In opposed anvil-type device the pressure mechanism is based on the compression and confinement of a solid medium between special working surfaces of a pair of opposed-anvil type dies (Fig. 1 Right panel). The temperature was determined by the power dissipated in a graphite heater calibrated in separate runs. The pressure was determined by controlling the applied force. High pressure synthesis products were characterized by X-ray diffractometry. Low-field magnetization data were collected by using a superconducting quantum interference device magnetometer to check superconductivity of the high pressure samples. Morphology of the crystal grains (SE images) and the chemical composition were measured using a scanning electron microscope (SEM-Cam Scan 44) with an energy dispersive X-ray analyzer (EDX-Edax Phoenix). The μSR experiments have been done at the πM3 and πE1 beam lines of the Paul Scherrer Institute (Villigen, Switzerland). Single crystals were studied by torque magnetometry using highly sensitive piezoresistive torque sensors our home design [5].

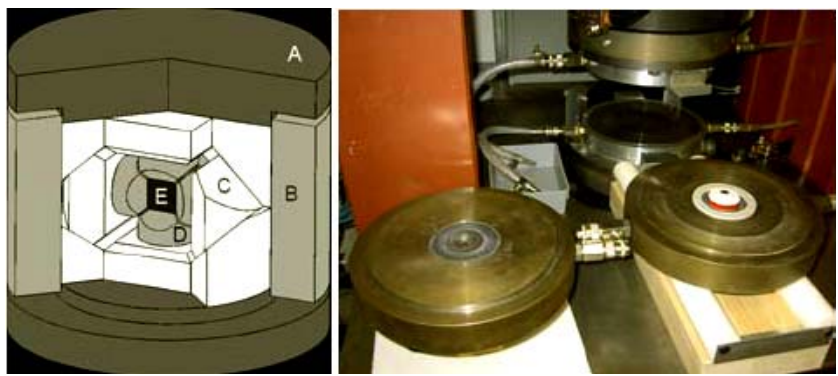

**Fig. 1.** Left panel-Cubic anvil-type high pressure device: A-base plate; B-load ring; C-steel parts; D-tungsten carbide pistons; E-sample volume. Maximum working pressure is 35 kbar and temperature 2200 °C. Right panel-Opposed anvils-type high pressure device. Pressure medium is lithographic stone. Maximum working pressure is 77 kbar and temperature 1700 °C.

## III. RESULTS AND DISCUSSION

Single crystals growth of $Ca_{2-x}Na_xCuO_2Cl_2$ is a technically complex task requiring a combination of high pressure and high temperature. Figure 2 summarizes the results of high-pressure growth experiments. Since our cubic anvil-type apparatus has the maximum working pressure of 35 kbar opposed-anvil type apparatus was used for experiments at higher pressures. Our earlier experiments [4] at 45 kbar show that the Na content depends not only on the synthesis pressure, but also on the reaction temperature and time. As we can see in Fig. 2, when the sample was compressed at 45 kbar and the temperature was slowly ramped from 1500 to 1300 °C within 5 h, the product showed $T_{c,on}$=28.0 K. Similar $T_c$ was observed when the samples were synthesized at 55 kbar. In this study further studies were performed in order to optimize the experimental conditions suitable for growing larger single crystals. In a typical run, a pressure of 30-55 kbar is applied at room temperature. While keeping pressure constant, the temperature is ramped up within 1.5 h to the maximum value of 1250-1700 °C, and is kept for 10-30 min and then slowly (-5-15 °C/h) cooled to 1050-1100 °C crossing the melting point of NaCl (see the insert on Fig. 2) and finally cooled to room temperature in 2-3 h.

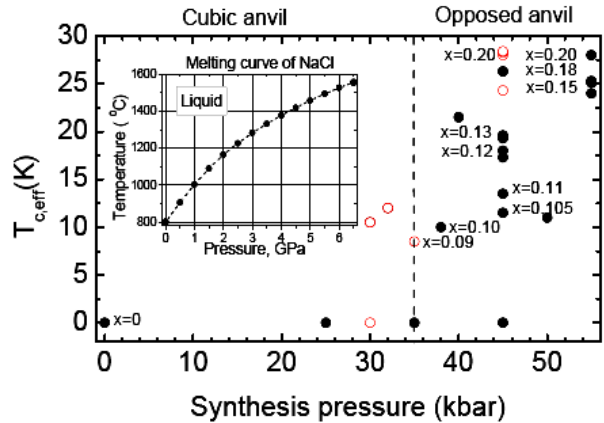

**Fig. 2.** The Na content ($x$) of $Ca_{2-x}Na_xCuO_2Cl_2$ was controlled by the synthesis pressure and temperature. As Na content increases $T_c$ changes systematically. Closed circles - polycrystalline samples, open circles-samples with collection of single crystals. Insert - melting curve of NaCl up to 65 kbar adapted from Ref. [6].

The temperature gradient, which is another important parameter, was fixed by placing the samples at the same position in the furnace. Our cubic-anvil high-pressure apparatus is much more favourable for crystal growth of Na-CCOC, because it has a higher temperature gradient than the opposed-anvil device. With the aim to further lower the synthesis pressure, we performed search experiments at pressures not higher than 35 kbar. The Na content depends on both the synthesis pressure and temperature. However, optimally doped Na-CCOC compounds can not be synthesized below 45 kbar. Figure 3 shows the normalized diamagnetic signal for a series of $Ca_{2-x}Na_xCuO_2Cl_2$ samples with various Na content. The biggest superconducting volume fraction ($\approx$100 %) and highest $T_{c,on}$=29 K were observed at the synthesis pressure of 45 kbar starting from nominal composition $Ca_2CuO_2Cl_2+0.2NaClO_4+0.4NaCl$. Although it is not easy to control the conditions with high reproducibility, general conclusions are following: (i) the size of grown crystals depends on the sample volume, (ii) the temperature gradient along the heater is crucial for crystal growth, (iii) when the growth temperature was higher than melting point of NaCl, the size of $Ca_{2-x}Na_xCuO_2Cl_2$ crystals drastically increased, (iv) prolonged cooling time improved the size and surface quality of single crystals, (v) starting composition require further optimization. After crushing the lump in an Ar-filled glove box, a large number of crystals of sizes of hundreds μm were found. The photographs of fractured samples of $Ca_{2-x}Na_xCuO_2Cl_2$ taken with an optical microscope are shown in Fig. 4. The morphology of $Ca_{2-x}Na_xCuO_2Cl_2$ crystals depends on the local sodium supersaturation which in turn depends on the local temperature distribution. The size of the as-grown crystals was bigger on outer wall of the cylindrical crucible (crystals marked by white contour lines on Fig. 4).

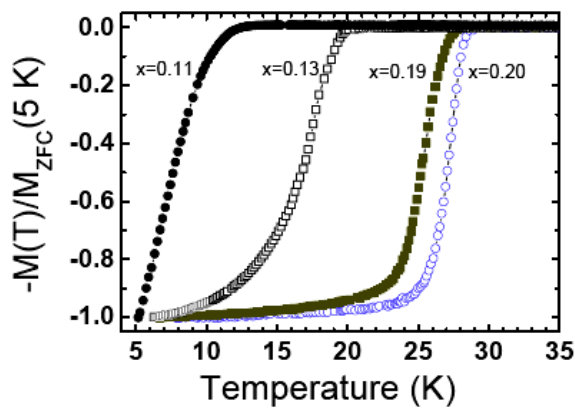

**Fig. 3.** Normalized diamagnetic signal for a series of $Ca_{2-x}Na_xCuO_2Cl_2$ samples with various Na content ($x$) grown at high pressure. The highest $T_{c,on}$=29.0 K was observed for optimally doped composition $Ca_{1.8}Na_{0.2}CuO_2Cl_2$.

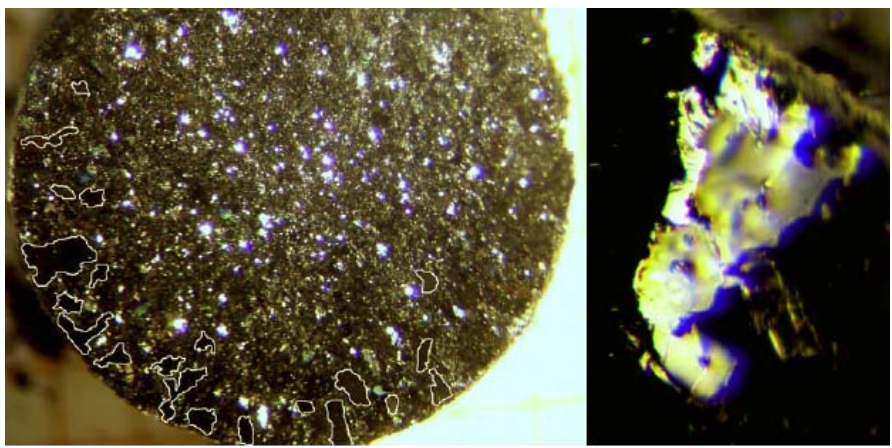

**Fig. 4.** Photographs of fractured samples of $Ca_{2-x}Na_xCuO_2Cl_2$ taken with an optical microscope. Many crystals of several hundreds μm sizes are seen (left). Right picture shows magnification of an individual single crystal.

The extracted $Ca_{2-x}Na_xCuO_2Cl_2$ crystals exhibit platelet shape and black colour (Fig. 5). As we can see from the SE images on the Fig. 5 the as-grown crystals contain cracks due to interaction between crystals growing from many nucleation sites. In the EDX analysis we have concentrated on the platelet-shaped crystals with clean surface. Our comparative studies did not show significant differences between platelets. The semi-quantitative evaluation of the sodium content was obtained from the EDX data and these values correlate well with those obtained by analysing of the *c*-axis lattice parameter dependence as a function of the Na content [1,3,4]. Reconstructed reciprocal layer *hk0* perpendicular to the *c* direction of almost optimally doped $Ca_{1.82}Na_{0.18}CuO_2Cl_2$ is shown in Fig. 6. Single crystal analysis confirmed tetragonal structure with lattice constants $a$=3.8457(11) Å and $c$=15.176(7) Å.

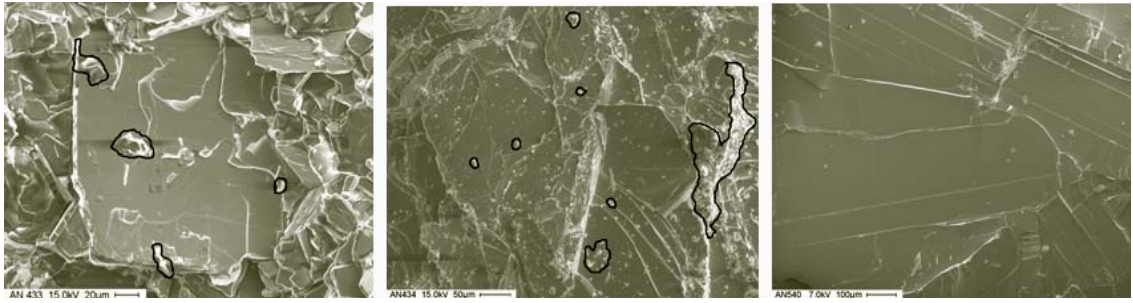

**Fig. 5.** From left to right. SE images of $Ca_{2-x}Na_xCuO_2Cl_2$ with $x \approx 0.18$ and $T_c=27$ K; $x \approx 0.11$ and $T_c=13$ K; $x \approx 0.06$, not superconductor. Some rests of solidified flux can be seen on the surface of crystals (marked by black contour lines).

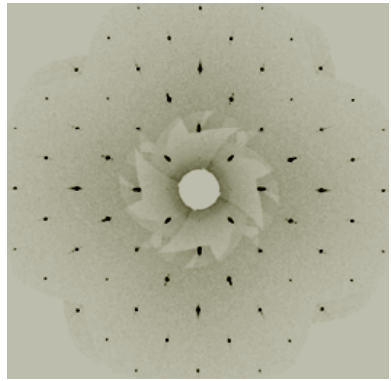

**Fig. 6.** Reconstructed reciprocal layer $hk0$ of the $Ca_{1.82}Na_{0.18}CuO_2Cl_2$ single crystal perpendicular to the $c$ direction.

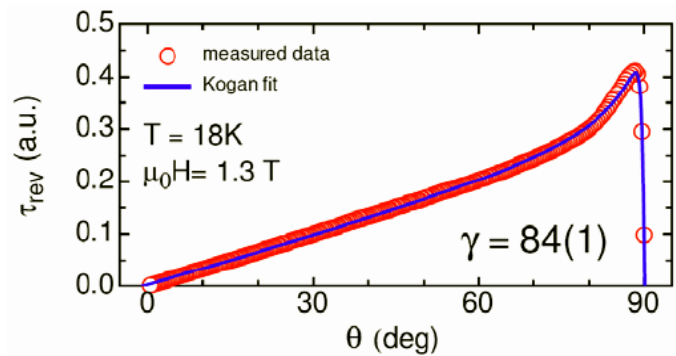

**Fig. 7.** Measured torque in arbitrary units of $Ca_{1.82}Na_{0.18}CuO_2Cl_2$ single crystal as a function of the angle between the applied magnetic field of 1.3 T and the $c$-axis of the crystal.

Physical properties of $Ca_{2-x}Na_xCuO_2Cl_2$ samples were studied by means of torque magnetometry and muon-spin rotation techniques. Torque studies of $Ca_{1.82}Na_{0.18}CuO_2Cl_2$ single crystals have been performed by using highly sensitive piezoresistive torque sensors. During the experiment, magnetic field is turned according to the crystallographic axes of the sample so that the full angular torque of the sample can be investigated using the shaking technique [7]. The anisotropy coefficient $\gamma=\lambda_c/\lambda_{ab}$ of the $Ca_{1.82}Na_{0.18}CuO_2Cl_2$ single crystals was determined to be $\gamma=84$ by fitting Kogans relation [8] to the measured torque. The result is in agreement with the work of Kim et al. [9]. They estimated 50-800 as a possible range for gamma. The critical temperature was estimated to lie at 27 K and $B_{c2}(0)$ is expected to be between 15 T and 20 T. Some sample-dependent deviations of the anisotropy parameter have been observed. They are due to crystal imperfection such as the slightly varying angle (by a few degrees) between *c*-axis and the *ab*-plane, leading to a smearing of the experimental curve close to the *ab*-plane. Therefore, only the crystals, which showed a minimum of this artificial smearing, where analyzed in order to determine the anisotropy parameter as precisely as possible. Our results implies that the anisotropy of $Ca_{1.82}Na_{0.18}CuO_2Cl_2$ is much more enhanced compared to the structurally related $La_{1.82}Sr_{0.18}CuO_4$, where for same doping level $\gamma\approx11$ [10].

In order to check the role of the apical oxygen for high-temperature superconductivity, we performed µSR and low-field magnetization studies of the in-plane magnetic field penetration depth $\lambda_{ab}$ for $Ca_{2-x}Na_xCuO_2Cl_2$ samples with Na content $x\approx0.11, 0.12, 0.15, 0.18$, and $0.19$ [11]. In µSR experiments $\lambda_{ab}$ was obtained from the measured muon-spin depolarization rate in a superconductor in a mixed state. For the polycrystalline $Ca_{2-x}Na_xCuO_2Cl_2$ superconductors, the envelope of the muon precession signal has approximately a Gaussian form $exp(-\sigma^2 t^2/2)$ and the depolarization rate $\sigma^2=\sigma_{sc}^2+\sigma_n^2$, where $\sigma_{sc}$ is the spin depolarization rate due to the perturbation of the vortex lattice, and $\sigma_n\approx0.1$ µs$^{-1}$ is the spin depolarization rate due to nuclear dipole fields. From the obtained values $\sigma_{sc}(0)$ the absolute values of $\lambda_{ab}$ were determined as:

$$\lambda_{ab}(\text{nm}) = 224/\sqrt{\sigma_{sc}(\mu^{-1})} \qquad (1)$$

In the low-field magnetization experiments $\lambda_{ab}$ was determined by using the procedure described by Kanigel et al. [12]. It was shown that for HTS powder samples, shaped in a cylindrical container having a diameter much smaller than its length, $\lambda_{ab}(0)$ can be obtained from the so-called intrinsic susceptibility $\chi^{int}(0)=M_{FC}(0)/M_{id}$ ($M_{id}$ is the magnetization of an ideal diamagnet) according to the relation $\chi^{int}(0)\propto\lambda_{ab}^{-2}(0)$. The results of $\lambda_{ab}$ studies are summarized in Table 1.

**Table 1.** Summary of the $\lambda(T)$ study of $Ca_{2-x}Na_xCuO_2Cl_2$ (see text for details). [a]Normalized to $\lambda_{ab}(0)=317(19)$ nm for $x\approx0.18$ sample in µSR experiment.

| Method | $x$ | $T_c$ (K) | $\sigma_{sc}(0)$ (µs$^{-1}$) | $\chi^{int}(0)$ | $\lambda_{ab}(0)$ (nm) |
|---|---|---|---|---|---|
| µSR | 0.12 | 18.5(2) | 0.33(2) | - | 390(24) |
| µSR | 0.18 | 25.6(2) | 0.50(3) | - | 317(19) |
| $\chi^{int}(0)$ | 0.11 | 15.14(4) | - | 0.171 | 430(26) |
| $\chi^{int}(0)$ | 0.12 | 18.40(3) | - | 0.192 | 406(24) |
| $\chi^{int}(0)$ | 0.15 | 23.45(3) | - | 0.307 | 321(19) |
| $\chi^{int}(0)$ | 0.18 | 27.11(3) | - | 0.315 | 317(19)[a] |
| $\chi^{int}(0)$ | 0.19 | 27.70(3) | - | 0.316 | 316(19) |

The absolute value of the in-plane magnetic penetration depth at $T=0$ was found to increase with decreasing doping from $\lambda_{ab}(0)=316(19)$ nm for $x\approx0.19$ sample to $\lambda_{ab}(0)=430(26)$ nm for the $x\approx0.11$ one. Comparison of the superfluid density $\rho_s\propto\lambda_{ab}^{-2}(0)\propto\sigma_{sc}(0)$ of Na-CCOC with that for the structurally related La214 compound [13] reveals that for the same doping level, $\rho_s$ in Na-CCOC is more than a factor 2 smaller than in La214. The reason for this is very likely due to a substantial decrease of the amount of holes on the apical sites in Na-CCOC.

## IV. SUMMARY

Polycrystalline samples and submillimeter size single crystals of Na-doped $Ca_2CuO_2Cl_2$ have been synthesized under high pressure. A series of experiments showed that the Na content depends not only on the pressure during the synthesis but also on the synthesis temperature and time. From a comparison of the Na-CCOC data with those of structurally related La214 cuprate superconductors we concluded that chlorine at the apical site is less effective that oxygen in supplying charge carriers to the $CuO_2$ plans. As a result, the coupling between the $CuO_2$ planes is weakened, the transition temperature $T_c$ is reduced and the anisotropic nature is enhanced.

## ACKNOWLEDGEMENTS

This work was supported by the NCCR program MaNEP. The authors are grateful to Electron Microscopy ETH Zurich (EMEZ) for SEM and EDX analysis and R. Puzniak for helpful discussion of the torque data.

## REFERENCES

[1] Z. Hiroi, N. Kobayashi and M. Takano, *Nature* **371** 139 (1994)
[2] T. Hanaguri, C. Lupien, Y. Kohsaka, D.H. Lee, M. Azuma, M. Takano, H. Takagi and J.C. Davis, *Nature* **430** 1001 (2004)
[3] Y. Kohsaka, M. Azuma, I. Yamada, T. Sasagawa, T. Hanaguri, M. Takano and H. Takagi, *J. Am. Chem. Soc.* **124** 12275 (2002)
[4] N.D. Zhigadlo and J. Karpinski, *Physica C* **460-462** 372 (2007)
[5] S. Kohout, J. Roos and H. Keller, *Rev. Sci. Inst.* **78** 013903 (2007)
[6] J. Akella, S.N. Vaidya and G.C. Kennedy, *Phys. Rev.* **185** 1135 (1969)
[7] M. Willemin, C. Rossel, J. Hofer, H. Keller, A. Erb and E. Walker, *Phys. Rev. B* **58** 5940R (1998)
[8] V.G. Kogan, M.M. Fang and Sreeparna Mitra, *Phys. Rev. B* **38** 11958R (1988)
[9] K.H. Kim, H.J. Kim, J.D. Kim, H.G. Lee and S.I. Lee, *Phys. Rev. B* **72** 224510 (2005)
[10] S. Kohout, T. Schneider, J. Roos, H. Keller, T. Sasagawa and H. Takagi, *Phys. Rev. B* **76** 064513 (2007)
[11] R. Khasanov, N.D. Zhigadlo, J. Karpinski and H. Keller, *Phys. Rev. B* **76** 094505 (2007)
[12] A. Kanigel, A. Keren, A. Knizhnik and O. Shafir, *Phys Rev B* **71** 224511 (2005)
[13] J.L. Tallon, J.W. Loram, J.R. Cooper, C. Panagopolous and C. Bernhard, *Phys. Rev. B* **68** 180501R (2003)